  \def\om{\omega}

\def\imo{i}

\def\be{\begin{equation}}
\def\ee{\end{equation}}
\def\bea{\begin{eqnarray}}
\def\eea{\end{eqnarray}}

\documentclass[11pt]{article}
\usepackage{graphicx}
\usepackage{longtable}
\begin{document}
\centerline{\bf \Large Influence of the back reaction of the Hawking}
\centerline{\bf \Large radiation upon black hole quasinormal modes}
\bigskip
\renewcommand{\thefootnote}{\fnsymbol{footnote}}
\setcounter{footnote}1
\centerline{\bf R. A. Konoplya\footnote{E-mail: konoplya$\mathrm{_{-}}$roma@yahoo.com}}
\renewcommand{\thefootnote}{\arabic{footnote}} \setcounter{footnote}0
\bigskip

\centerline{Department of Physics, Dniepropetrovsk National University}
\centerline{St. Naukova 13, Dniepropetrovsk 49050, Ukraine}
\vskip3mm
\bigskip
\nopagebreak

\begin{abstract}
We consider the BTZ black hole surrounded by the conformal scalar
field. Within general relativity, the resonant
\emph{quasinormal} (QN) modes dominate in the response of a black
hole to external perturbations. At the same time, the metric of an
evaporating black hole is affected by the Hawking radiation. We
estimate the shift in the quasinormal spectrum of the BTZ black
hole stipulated by the back reaction of the Hawking radiation. For
the case of the 2+1 dimensional black hole the corrected (by $\sim
\hbar$) metric is an \emph{exact} solution [C.Martines, J.Zanelli
(1997)]. In addition, in this case quantum corrections come only
from matter fields and no from graviton loops, that is, one can
solve the problem of influence of the back reaction upon the QN
ringing self-consistently. The dominant contribution to the
corrections to the QNMs is simply a shift  of $\omega^{2}$
proportional to $-(\frac{\Lambda}{M})^{3/2} (4 L^{2} +M) \hbar$.
It is negligible for large black holes but essential for small
ones, giving rise to considerable increasing of the quality
factor. Thus, the small evaporating black hole is expected to be
much better oscillator than a large one.
\end{abstract}

\twocolumn
In classical regime a black hole does not emit anything. It is
characterized by its three parameters: mass, charge, and angular
momentum. When perturbing a black hole the background geometry
undergoes damping oscillations dominated at late times by the so
called \emph{quasinormal} modes. They are of a great importance
because they depend upon the above parameters of a black hole only
and not on the way of excitation.  Thus these modes represent the
characteristic resonance spectrum of a black hole response (see
\cite{kokkotas-review} for a review). The QN modes have gained
considerable interest owing to their interpretation in ADS/CFT
correspondence
\cite{1h}, \cite{1k},  \cite{1ckl}, \cite{1starina}, \cite{cl},   and Loop Quantum
Gravity (see for instance
\cite{2} and references
therein). Recent investigation of black holes within brane models
stimulated the calculation of QN modes of different higher
dimensional black holes (see \cite{3} and references therein). In
addition, the QN radiation of dilaton black holes have been
recently studied in
\cite{dilatonQNM}

At the same time a black hole radiates energy with thermal
spectrum, when taking into account the effect of quantized fields
near the black hole. Thus a black hole can exist in a thermal
equilibrium with a heat bath composed of quantum fields
interacting with the black hole geometry \cite{Hawking-75CMP}. In
four dimensions the back reaction problem is solved usually as
follows: one considers the expectation value of the renormalized
(approximate) stress-energy tensor in appropriate "vacuum" state
\cite{Page-T} as the source in the Einstein equations and solves
these equations self- consistently for the metric
\cite{York-85}.

In four dimensions the corrected metric diverges at large r, and
in order to restrict cumulative effect from the corrected
geometry, one need to put a shell outside of which the geometry is
"uncorrected". As a result the final metric inside the shell
contains a constant, which is determined by boundary conditions at
the shell. The latter is assumed to be posed at some fixed
distance from the event horizon. The picture significantly depends
upon this boundary condition at the shell. Thus the two models are
generally accepted. First, when one specifies the total energy of
the system  at the shell. That is micro-canonical ensemble. The
second choice, a canonical ensemble, is to fix the temperature at
the shell (see \cite{4} for recent references). If one would like
to find the QN modes of such "corrected" metric one have to deal
with a step-function (or delta-function) in the corresponding
effective potential, at the radius of the shell $r_{0}$.  This
delta- function would crucially change the eigenvalues to be
determined. The search of QN modes for such a "dirty" black hole
should be done in a model independent way, in order, for example,
that the found modes would not depend upon $r_{0}$. What is even
more important in four dimensions, that if taking into
consideration corrections from quantized fields of order $\hbar$,
one must include corrections of the same order coming from quantum
gravity.

Fortunately, in 2+1 dimensions the situation is much easier. First
of all, the 2+1 gravity has no propagating degrees of freedom and
at each point the Riemann tensor is completely determined by the
matter source there. A quantum gravity in 2+1 dimensions is
renormalizable and finite \cite{Witten-88}. Thus the only
radiative corrections to the geometry are coming from quantum
excitation of the matter fields, and, the perturbative expansion
receives no corrections from graviton loops
\cite{Witten-88}. At the same time there is a useful black hole
solution in three dimensions with negative cosmological constant,
the Banados-Teitelboim-Zanelli (BTZ) black hole \cite{BTZ}.


The QN behavior of asymptotically anti-de Sitter (ADS) black holes
\cite{1h},
\cite{1starina} crucially depends upon the
black hole size relative to the ADS radius: for large BHs the QN
modes are proportional to the radius of a black hole
\cite{1h}, while for small black holes they approach the modes of
the empty ADS space-time \cite{1k}. The ADS space-time forms an
effective confining box, and the potential diverges at spacial
infinity. The case of conformal scalar field is different since
the potential approaches a constant at infinity. That is why the
QN behavior of conformal scalar field is different from that of
the "ordinary" minimally coupled field studied in \cite{1h},
\cite{1k},
\cite{1ckl}. The QN modes of the BTZ black hole were calculated
for conformal scalar field in \cite{Mann}, and for non-conformal
scalar, electromagnetic and dirac fields in \cite{cl}.


Consider the system consisting of the BTZ black hole and the
conformal scalar field surrounding it. Let us find out what will
happen with QN modes which govern the decay of this conformal
scalar field if taking into account the back reaction of quantum
radiation of the same field upon the surrounding geometry. For
this case and transparent boundary conditions at infinity the
stress-energy tensor $<T_{\mu \nu}>$ was calculated in
\cite{Steif}. The \textit{O($\hbar$)} correction to the black hole
geometry due to the radiative conformal field is governed by the
semiclassical equations:
\begin{equation}
G_{\mu \nu} + \Lambda g_{\mu \nu }  = \kappa <T_{\mu \nu}>.
\end{equation}
An \emph{exact} solution of these equations was found by Martines
and Zanelli in \cite{Martines-Zanelli-BR}:
$$ d s^{2} = - f(r) d t^{2} + f^{-1}(r)d r^{2} + r^{2} d \theta^{2}, $$
\begin{equation}
f(r)=\left(r^{2} \Lambda - M - \frac{2 l_{p} F(M)}{r}\right),
\end{equation}
Here $F(M)$ is determined in the following way \cite{Steif}:
$$ F(M) = $$
\begin{equation}
\frac{M^{3/2}}{2 \sqrt{2}}  \sum_{n=1}^{\infty} e^{-i n
\delta} \frac{Cosh[2 \pi n \sqrt{M}] +3}{(Cosh[2 \pi n \sqrt{M}] -1)^{3/2}},
\end{equation}
where $\delta$ is an arbitrary phase.

We used $G=1/8$, $M$ can be associated with the black hole mass
\cite{zaslavskiiBTZ1}, $l_{p} =
\hbar/8$ is the Plank mass in three dimensions, the Plank mass
$m_{p} = \hbar/l_{p} = 8$ is independent of $\hbar$. The series
(2) converges exponentially for any $M > 0$. For $M \gg 1$ the
first term dominates the series $F(M) \sim e^{-\sqrt{-\pi M}}
\rightarrow 0$ and the BTZ black hole is recovered.
The metric (2) is an exact solution of the back reaction problem
for the one-loop effective energy momentum tensor of a scalar
field conformally coupled to gravity.  Formally the metric (2)
coincides with an exact solution for the BTZ black hole "dressed"
by conformal scalar field \cite{Martines-Zanelli}.  

Next, we shall consider the corrected metric $(2)$ as a background
for conformal scalar field and find the corresponding QN spectrum.
The conformally coupled scalar wave equation has the form:
\begin{equation}
\nabla^{2} \Phi(t, r, \theta)  = \frac{1}{8} R \Phi(t, r, \theta).
\end{equation}
After the change of the wave function $\Phi = \Psi/\sqrt{r}$, and
the radial coordinate $d r^{*} = dr/f(r)$, and, separation of
angular and time coordinates $t$ ($\Psi
\sim e^{i
\omega t}$) and $\theta$ ($\Psi \sim e^{i L \theta}$) one comes to
the wave equation:
\begin{equation}
\left(\frac{d^2}{dr^{*2}} + \om^2 - V\right)\Psi(r^*) = 0,
\end{equation}
where the potential $V$ has the following form
\begin{equation}
V = \left(\frac{M + 4 L^{2}}{4 r^{2}} - \frac{3 l_{p} F(M)}{2
r^{3}}\right) f(r).
\end{equation}
In the considered range of parameter $M$, this potential as a
function of $r^{*}$ approaches its maximum at $r^{*} = 0$ (spacial
infinity) and goes to zero at $r^{*} = -
\infty$ (horizon) without any barriers near the black hole horizon
(as it takes place for conformal scalar field around SAdS black
hole \cite{Mann}). Thus the effective potential of the quantum
corrected BTZ black hole has the same features as that of the
"pure" BTZ black hole.

In asymptotically flat space-time  the QN modes are determined as
the eigenvalues $\omega$ such that, under the choice of the
positive sign of the real part of $\omega$, QNMs satisfy the
following boundary conditions
\begin{equation}
\Psi(r^*) \sim C_\pm \exp(\pm\imo\om r^*), \qquad r^*
\longrightarrow  \pm\infty,
\end{equation}
corresponding to purely in-going waves at the event horizon and
purely out-going waves at spacial infinity. In our case the
space-time is asymptotically anti-de Sitter and the appropriate
boundary condition at spacial infinity is the Dirichlet one
\cite{Avis-78}, while at the horizon it is, certainly, the
requirement of purely in-going waves.

From here and on, in order to find the dominant contribution to
the QN spectrum from  \textit{O($\hbar$)} correction to the BTZ
space-time, we shall neglect the order of $l_{p}$ higher than
first. Thus inverting the $r$ coordinate as a function of $r^{*}$
we find up to the first order of $l_{p}$:
$$ r(r^{*}) = \sqrt{\frac{M}{\Lambda}} \frac{1+
\alpha^{2}}{1-\alpha^{2}}+  $$
\begin{equation}
\frac{F(M)}{2 M}
\left(2- \frac{16 (\ln{\alpha}+ \ln{(\alpha+1))}}{\alpha^{2} +\alpha^{-2}
+2}\right) l_{p} + {\it O(l_{p})},
\end{equation}
where $\alpha = e^{\sqrt{\Lambda M} r^{*}}$. The $r^{*}$ goes from
$-\infty$ to $0$ as $r$ goes from the event horizon to infinity.
Thus the value $e^{\sqrt{\Lambda M} r^{*}}$ is always less than
$1$ and we can, following the paper \cite{Mann}, expand the
effective potential into series of powers of $\alpha$. The first
term, as it was shown in \cite{Mann}, gives the dominant QN
behavior with good accuracy. Wishing to estimate {\it dominant}
contribution to the shift of the QN spectrum we shall be
restricted here by considering corrections greater than ${\it
O(l_{p}^{2}, \alpha^{4},
\alpha^{2} l_{p})}$. For large black holes it is understood that
next terms in $\alpha^{2}$ are more important than even first
correction $\sim \hbar$. Yet, in the regime of small black hole,
where the back reaction is significant, the higher order
corrections in $\alpha^{2}$ is less important than $\sim
l_{p}$-corrections. Thus, the approximated potential we shall
investigate, has the form:
$$ V(r^{*}) = \frac{(4 L^{2} +M) F(M)
}{(\frac{M}{\Lambda})^{3/2}} l_{p} +$$
\begin{equation}
 + (4 L^{2} +M) \Lambda e^{2
\sqrt{\Lambda M} r^{*}} +{\it O(l_{p}^{2}, \alpha^{4}, \alpha^{2} l_{p})}.
\end{equation}
The QN modes for "uncorrected" potential $V_{0}(r^{*}) = V_{0}
e^{2 \sqrt{\Lambda M} r^{*}}$, $V_{0} = (4 L^{2} +M) |\Lambda | $,
were calculated in
\cite{Mann}. Comparison of the results obtained through $V_{0}(r^{*})$
with higher order corrections in $\alpha$ shows that the dominant
behavior is stipulated by this approximated potential
$V_{0}(r^{*})$
\cite{Mann}. In fact, the above potential  $V$ given by the formula
$(9)$ differs from $V_{0}(r^{*})$ only by a constant shift
$\frac{(4 L^{2} +M) F(M) }{(\frac{M}{\Lambda})^{3/2}} l_{p}$,
which simply can be thought of as a shift of $\omega^{2}$.
Therefore, an exact solution of the wave equation $(5)$ will have
the similar form as that obtained in \cite{Mann}. Namely, the
Green function $G(r^{*},
\xi;
\tilde{\omega})$, satisfying the wave equation
\begin{equation}
\left(\frac{d^2}{dr^{*2}} + \tilde{\om}^2 - V\right) \tilde{G}(r^{*}, \xi; \tilde{\omega}) = -\delta(r^{*} - \xi),
\end{equation}
has the form:
$$ G(r^{*}, \xi< r^{*}; \tilde{\omega}) =$$
\begin{equation}
\frac{I_{\nu} (z(\xi)) [I_{\nu}
(Z_{0}) K_{\nu} (z(r^{*}))- K_{\nu} (Z_{0}) I_{\nu}
(z(r^{*}))]}{\sqrt{\Lambda M} I_{\nu} (Z_{0})}.
\end{equation}
Here
\begin{equation}
\tilde{\omega^{2}} = \omega^{2} - \frac{(4 L^{2} +M)
F(M)}{(\frac{M}{\Lambda})^{3/2}} l_{p},
\end{equation}
$$\quad \nu = - i
\tilde{\omega}/\sqrt{\Lambda M}, \quad z= \sqrt{\frac{V_{0}}{\Lambda M}}
\alpha, \quad Z_{0} = \sqrt{\frac{V_{0}}{\Lambda M}}.$$
The QN modes are the poles of this Green function and thereby are
zeros of the modified Bessel function
\begin{equation}
I_{\nu} (Z_{0}) = 0
\end{equation}
We see (Fig.1, 2, 3) that this shift, being negligible for large
black holes, becomes significant for small black holes and gives
rise to increasing of the real oscillation frequency  and to
decreasing of the damping rate in this regime. Therefore the
quality factor, which is proportional to
$\mid\omega_{Re}\mid/\mid\omega_{Im}\mid$, is increasing
considerably when one goes over to considering of  smaller mass of
the black hole and, at the same time, including the back reaction
of the Hawking radiation. From this, one can conclude that the
small evaporating black hole is expected to be much better
oscillator than a large one. Remember, that the quality factor of
the large Schwarzschild black hole is of order $L$ at the
fundamental overtone which is, for instance, roughly $10^{6}$
times is less than that of an atom. That is, the large black hole
is a very poor oscillator
\cite{Mashhoon-cqg}. Note also, that for very small mass, next
corrections in $\hbar$ should be considered in the semiclassical
equations.

The QN frequencies shown in figures 1, 2, 3 are found under the
Dirichlet boundary conditions as closest to the $\omega_{Re}$-axis
poles of the modified Bessel function. Nevertheless, the shift
given by the formula (12) does not depend upon the boundary
conditions to be chosen. The dependence on $L$ of the QNMs is
demonstrated on Fig.4. We see that both $\omega_{Re}$ and
$\omega_{Re}$ are roughly proportional to $L$.

The influence of the back reaction on higher overtones is simply
the above shift  given by (12) and certainly is negligible for
modes with huge imaginary part. The higher overtones can be found
by extensive numerical search of the zeros of the modified Bessel
function. The higher overtone behavior strongly depends upon the
value of $z$: while first several overtones have both
non-vanishing real and imaginary parts, the higher ones have tiny
real parts, and the more $z$, the greater the number of modes with
non-vanishing real part.
Asymptotically, for highly damping modes, governed by an
approximated potential (9), one has
\begin{equation}
Re \tilde{\omega} \rightarrow 0, \quad Im \tilde{\omega}
\rightarrow  n + z - 1
\quad as \quad n \rightarrow \infty.
\end{equation}
Note, that this asymptotic regime comes very rapidly, i.e. it
takes place already at fifth overtone for $z=3$ and at somewhat
greater overtone number for greater $z$. This let us hope that the
same asymptotic behavior will take place when considering complete
effective potential with no approximations. This quick falling
into the asymptotic regime repeats the high overtone behavior of
non-conformal scalar field around ADS black hole
\cite{1ckl}.

Note that under the metric perturbations of the above mentioned
conformally dressed black hole \cite{Martines-Zanelli} there
appear the physically accepted growing gravitational modes if
imposing Dirichlet boundary conditions \cite{martinez-alone}. Even
though this indicates upon classical instability of the black
hole, the considered here spectral problem for the system,
consisting of the black hole and the conformal scalar field,
remains consistent since we are interested in study of decay of
the scalar field only and there is no coupling with gravitational
perturbations. For realistic 4-dimensional models such instability
would certainly "cut off" the motivation of study of the QN
spectrum. In three dimensions it is much more important that we
have consistent quantum corrected solution allowing to avoid
considering the problem in the realm of quantum gravity. After
all, the obtained shift of $\omega^{2}$ does not depend upon
boundary conditions which are very controversial in anti de Sitter
space-time
\cite{Avis-78,Mann,1starina,Moss-Norman}.

\textbf{Conclusion}.
We have estimated the dominant contribution to the back reaction
shift of the quasinormal modes for BTZ black hole surrounded by
conformal scalar field. It is interesting that the considering of
the effect of back reaction on the metric gives rise to the sharp
increasing of the quality factor of small black holes. This means
that a small black hole is a much better oscillator than a large
one, and, therefore, investigation of the resonance quasinormal
spectrum for such black holes should be important.


\begin{figure}
\begin{center}
\includegraphics{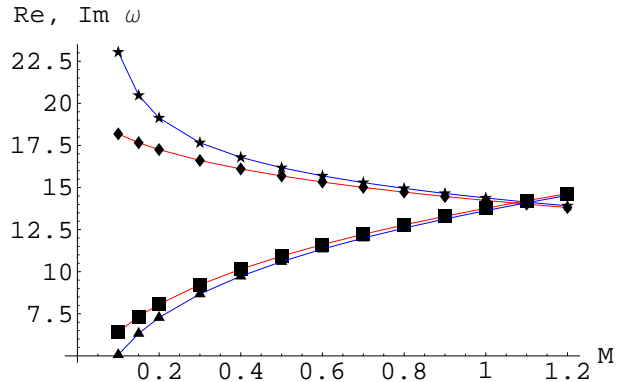}
\caption{$\omega_{Re}$ (top) and $\omega_{Im}$ (bottom) parts of $\omega$ for BTZ BH without back reaction (box, diamond)
and with back reaction (star, triangle) ($L=2$, $\Lambda=30$,
$\delta=0$, $n=0$) as a function of $M$}.
\label{1}
\end{center}
\end{figure}

\begin{figure}
\begin{center}
\includegraphics{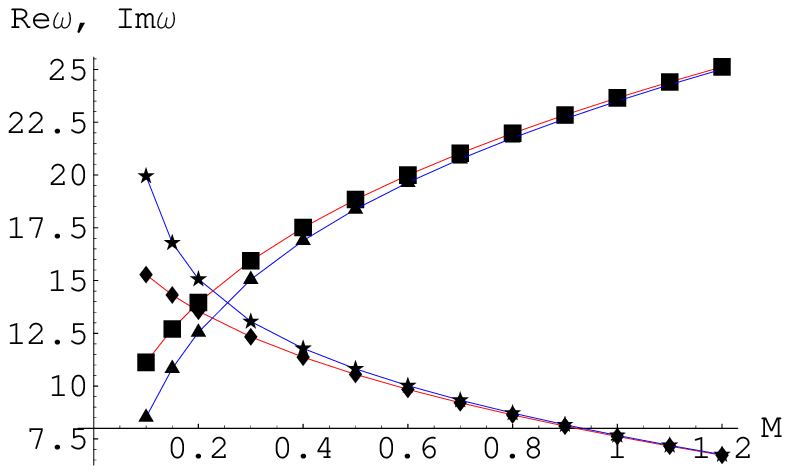}
\caption{$\omega_{Re}$ (top) and $\omega_{Im}$ (bottom) parts of $\omega$ for BTZ BH without back reaction (box, diamond)
and with back reaction (star, triangle) ($L=2$, $\Lambda=30$,
$\delta=0$, $n=1$) as a function of $M$}.
\label{1}
\end{center}
\end{figure}

\begin{figure}
\begin{center}
\includegraphics{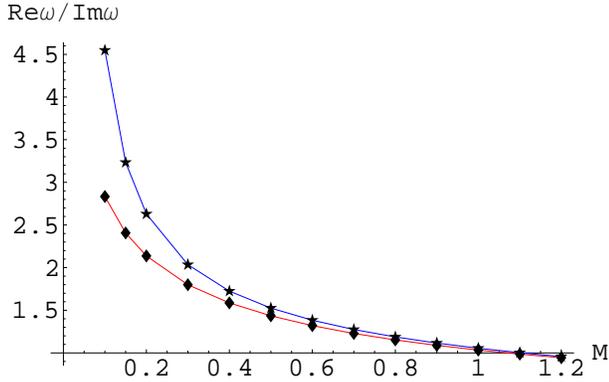}
\caption{The ratio $\omega_{Re}/\omega_{Im}$ for BTZ BH without back reaction (star)
and with back reaction (diamond) ($L=2$, $\Lambda=30$, $\delta=0$,
$n=0$) as a function of $M$}
\label{2}
\end{center}
\end{figure}



\begin{figure}
\begin{center}
\includegraphics{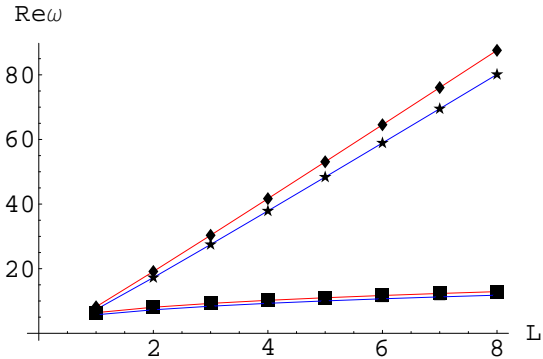}
\caption{$\omega_{Re}$ (top) and $\omega_{Im}$ (bottom) parts of QNm without (box, star) and with
(diamond, triangle) back reaction for different values of $L$,
($M=0.2$, $\delta=0$, $n=0$)}
\label{2}
\end{center}
\end{figure}


\begin{thebibliography}{99}
\bibitem{York-85} J.W.Jork, Jr., \texttt{Phys. Rev.} {\bf{D31}} 775 (1985)
(Note that the first, and, as far as we are aware, the single,
mention about the possible influence of the back reaction on QNMs
was done in this paper)
\end{thebibliography}
\end{document}